\documentclass[journal]{IEEEtran}

\usepackage{amsmath,amssymb}
\usepackage{booktabs}
\usepackage{array}
\usepackage{graphicx}
\usepackage{multirow}
\usepackage{algorithm}
\usepackage{algorithmic}
\usepackage{tikz}
\usetikzlibrary{arrows.meta,positioning,fit,shapes.multipart}
\usepackage{xurl}
\usepackage{cite}
\usepackage{balance}
\usepackage{microtype}
\newcommand{\sem}[1]{[\![#1]\!]}

\newcommand{\Kind}{\mathsf{Kind}}
\newcommand{\Base}{\mathsf{Base}}
\newcommand{\Beh}{\mathsf{Behavior}}
\newcommand{\BGroup}{\mathsf{BehaviorGroup}}
\newcommand{\PSet}{\mathsf{PropertySet}}
\newcommand{\Msg}{\mathsf{Message}}

\newcommand{\Trace}{\mathsf{Trace}}

\begin{document}

\title{From Token Interfaces to Token Semantics: A Formal Composition and Conformance Model for Implementation-Neutral Token Specifications}

\author{%
\begin{tabular}{c}
John deVadoss\\
co-Founder InterWork Alliance\\
Washington DC\\
johnd@ieee.org
\end{tabular}%
}

\makeatletter
\def\ps@plain{%
  \def\@oddhead{}%
  \def\@evenhead{}%
  \def\@oddfoot{\hfil\thepage\hfil}%
  \def\@evenfoot{\hfil\thepage\hfil}%
}
\def\ps@headings{%
  \def\@oddhead{}%
  \def\@evenhead{}%
  \def\@oddfoot{\hfil\thepage\hfil}%
  \def\@evenfoot{\hfil\thepage\hfil}%
}
\def\ps@IEEEtitlepagestyle{%
  \def\@oddhead{}%
  \def\@evenhead{}%
  \def\@oddfoot{\hfil\thepage\hfil}%
  \def\@evenfoot{\hfil\thepage\hfil}%
}
\makeatother

\pagestyle{plain}
\pagenumbering{arabic}

\maketitle
\thispagestyle{plain}

\begin{abstract}

Digital-token standards such as ERC-20, ERC-721, and ERC-1155 have been essential to blockchain adoption because they standardize callable interfaces. Interface standardization, however, does not fully specify token meaning. Two implementations may expose the same transfer function while encoding different assumptions about supply, divisibility, redemption, cancellation, evidence, governance, and lifecycle finality; conversely, two semantically equivalent tokens may be implemented on different ledgers and through different transaction models. This paper presents the InterWork Alliance Token Taxonomy Framework (TTF) as a typed semantic composition model for implementation-neutral token specifications. We formalize TTF artifacts, token formulas, behavior and property-set composition, well-formedness constraints, and a layered conformance model covering formula, artifact, message, state, and trace conformance. We further define a reference validation procedure and show how platform-neutral control messages can induce conformance obligations and implementation tests. The model is evaluated analytically through document-token, warehouse-receipt, and carbon/digital-MRV case studies, together with representative invalid compositions that interface standards alone do not expose. The analysis shows that token semantics can be specified, compared, validated, mapped, and governed independently of platform binding, providing a foundation for more reliable token interoperability across smart-contract platforms, permissioned ledgers, and shared-state systems.
\end{abstract}


\section{Introduction}

\IEEEPARstart{T}{okenization} has become one of the dominant abstractions in distributed ledger technology (DLT). A token may represent native digital value, a financial instrument, title to an off-chain asset, a warehouse receipt, a document, a ticket, a carbon-related claim, a software license, an educational credential, or a coordination primitive in a multiparty workflow. These artifacts are all called tokens, yet they differ in fungibility, divisibility, uniqueness, lifecycle, transfer restrictions, supply semantics, evidence requirements, governance assumptions, custody model, and legal or business meaning.

The industry has often handled this diversity through implementation standards. ERC-20 defines a standard interface for fungible tokens on Ethereum, including transfer, balance, allowance, and approval functions \cite{erc20}. ERC-721 defines a standard interface for non-fungible tokens whose individual assets are distinguishable and whose ownership must be tracked separately \cite{erc721}. ERC-1155 defines a multi-token interface in which one contract can manage any combination of fungible, non-fungible, and other token configurations \cite{erc1155}. These standards are foundational to wallets, marketplaces, developer reuse, and ecosystem scale.

Nevertheless, interface conformance is not semantic equivalence. A method signature alone does not tell a regulator whether a token is a transferable instrument, a non-transferable record, a claim on an off-chain asset, or a revocable workflow credential. A balance alone does not tell an auditor whether a burn operation denotes cancellation, retirement, redemption, or destruction of evidence. A non-fungible identifier alone does not tell a marketplace whether the object is a collectible, a ticket, a title document, or a regulated instrument. Similarly, two tokens may share the same business meaning while being implemented on different ledgers, in different languages, and under different transaction models.

The InterWork Alliance (IWA) Token Taxonomy Framework (TTF) addresses this problem at a different layer. The IWA describes TTF as an open-source resource library and common language for tokenization attributes, behaviors, and data properties, intended to bridge developers, businesses, and regulators \cite{gbbc_ttf}. The public TTF repository describes a framework composed of artifacts used to create token specifications, together with tools and an object model for token-design applications \cite{ttf_repo}. The taxonomy overview states that TTF breaks tokens into reusable parts - base token types, properties, and behaviors - and composes those parts into complete token definitions \cite{ttf_taxonomy}.

This paper argues that TTF is best understood not as a catalog of token names, but as a semantic intermediate representation for tokenization. A token definition should be precise enough to support implementation and certification, but neutral enough to survive mapping across Ethereum, Hyperledger Fabric, Corda, DAML, database-backed ledgers, and future shared-state systems. TTF provides such a layer through versioned artifacts, token formulas, property sets, behavior specifications, and implementation-neutral control messages.

\subsection{Contributions}

This paper makes four contributions.

\begin{enumerate}
\item It formalizes the TTF artifact model and token formula syntax as a typed composition language for implementation-neutral token definitions.
\item It defines a semantic interpretation of token formulas in terms of state spaces, control messages, invariants, transition relations, and observable traces.
\item It introduces a layered conformance model that distinguishes formula, artifact, message, state, and trace conformance, together with a reference validation procedure for detecting malformed or semantically inconsistent token specifications.
\item It applies the model to document, warehouse-receipt, and carbon/digital-MRV token patterns and analyzes the semantic gaps that remain invisible when tokens are classified only by implementation interfaces.
\end{enumerate}

The intended result is a stronger technical foundation for token interoperability: one in which semantic obligations can be specified, checked, mapped, and governed before code is deployed and after implementations are integrated.

\begin{figure*}[!t]
\centering
\resizebox{0.98\textwidth}{!}{%
\begin{tikzpicture}[
  node distance=1.4cm,
  box/.style={draw, rounded corners, align=center, minimum height=0.85cm, text width=3.1cm, font=\footnotesize},
  wide/.style={draw, rounded corners, align=center, minimum height=0.95cm, text width=4.2cm, font=\footnotesize},
  arr/.style={-{Latex[length=2.1mm]}, thick}
]
\node[box] (intent) {Business intent\\and domain requirements};
\node[box, right=of intent] (artifacts) {TTF artifact library\\bases, behaviors, property sets};
\node[box, right=of artifacts] (formula) {Token formula\\and token definition};
\node[box, right=of formula] (mapping) {Platform mapping\\Solidity, Fabric, Corda, DAML};
\node[box, right=of mapping] (impl) {Implementation\\contract, chaincode, state model};
\node[wide, below=1.25cm of formula] (conf) {Conformance obligations\\formula, artifact, message, state, trace};
\node[wide, below=1.25cm of mapping] (tests) {Generated checks and evidence\\control messages, invariants, traces};
\draw[arr] (intent) -- (artifacts);
\draw[arr] (artifacts) -- (formula);
\draw[arr] (formula) -- (mapping);
\draw[arr] (mapping) -- (impl);
\draw[arr] (formula) -- (conf);
\draw[arr] (conf) -- (tests);
\draw[arr] (tests) -- (impl);
\end{tikzpicture}%
}
\caption{TTF as a semantic intermediate layer. Business intent is decomposed into reusable artifacts, composed into a token definition, mapped to a platform, and checked through conformance obligations.}
\label{fig:architecture}
\end{figure*}
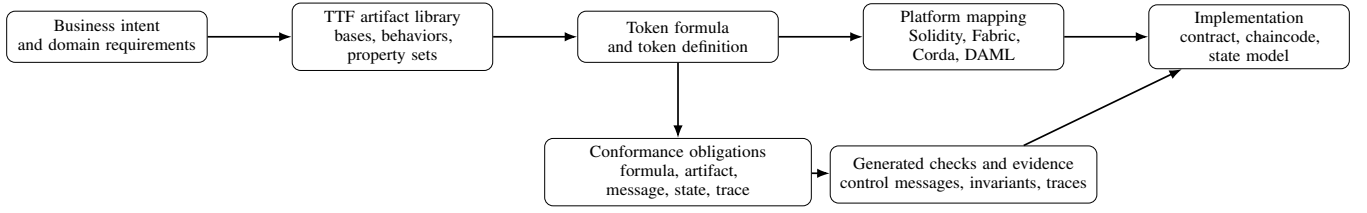

\section{Background and Related Work}

\subsection{Interface Standards and Token APIs}

ERC-20, ERC-721, and ERC-1155 are interface standards. They specify functions, events, and expected interaction patterns, thereby enabling wallets, exchanges, and applications to interoperate at the API boundary \cite{erc20,erc721,erc1155}. Their success also reveals their limitation. An interface standard defines how a token can be called on a platform, not all of what the token means in a business, accounting, regulatory, or evidence model.

NIST characterizes blockchains as tamper-evident and tamper-resistant digital ledgers implemented in distributed fashion and often without a central authority \cite{nist8202}. NIST's token design overview emphasizes that token systems involve multiple views, including token, wallet, transaction, user-interface, and protocol views \cite{nist8301}. TTF complements these views by focusing on the token-specification layer: the representation, behavior, data, and lifecycle commitments that should be understood before an implementation target is selected.

The Token Taxonomy Initiative was launched in 2019 as a blockchain-neutral effort to create common definitions, terminology, and specifications for tokens across platforms and industries \cite{eea_tti}. The TTF repository provides artifacts, an object model, documentation, control messages, and examples intended to support token-design applications \cite{ttf_repo,ttf_model,ttf_control}. This paper builds on those artifacts by giving them a formal composition and conformance interpretation.

\subsection{Semantic Interoperability and Model-Driven Engineering}

Semantic interoperability is the ability of independently developed systems to preserve meaning, not only syntax. Ontology engineering has long emphasized explicit specifications of conceptualizations and shared vocabularies \cite{gruber_ontology}. W3C ontology languages and related semantic-web work provide one path for machine-processable meaning, especially where concepts and relationships must be shared across organizations \cite{owl_overview}. TTF is narrower and more operational: it is not a general-purpose ontology, but a domain-specific composition model for token specifications.

Model-driven engineering similarly separates platform-independent models from platform-specific implementations. The Model Driven Architecture and subsequent model-driven engineering literature argue that models should be analyzable and transformable artifacts rather than documentation afterthoughts \cite{omg_mda,schmidt_mde}. Domain-specific languages make this separation concrete by giving a problem domain its own syntax and semantics \cite{fowler_dsl}. TTF can be read as a domain-specific language for token semantics: a formula indexes artifact definitions, and the expanded definition induces implementation and conformance obligations.

\subsection{Formal Methods, Smart Contracts, and Conformance}

Formal methods for smart contracts typically focus on proving or checking properties of deployed code or executable models. Prior work has explored formal verification of smart contracts, semantic frameworks for Ethereum, and model checking of state-transition systems \cite{bhargavan_smart,grishchenko_ethereum,clarke_model}. These methods are complementary to TTF. Verification asks whether an implementation satisfies a property; TTF helps define which semantic properties should be required in the first place.

Conformance testing has a long history in protocol standardization, where implementations are tested against externally specified behavior rather than judged only by internal design \cite{iso9646}. TTF control messages provide an analogous boundary for token implementations. A Solidity function, Fabric transaction, Corda command, or DAML exercise may realize the same semantic behavior through different native mechanisms; conformance should therefore be checked against token-level obligations, not merely platform-local APIs.

\section{Problem Statement and Requirements}

Let $I$ be an implementation interface and $S$ be the intended semantic specification of a token. Interface-first standardization assumes that agreement on $I$ is sufficient for interoperability. In practice, interoperability fails when distinct semantic specifications map to indistinguishable interfaces or when the same semantic specification maps to different platform interfaces:

\begin{equation}
S_1 \ne S_2 \quad \wedge \quad I(S_1) \equiv I(S_2),
\end{equation}

\begin{equation}
S_1 \equiv S_2 \quad \wedge \quad I(S_1) \ne I(S_2).
\end{equation}

The first case occurs when two tokens expose the same API but disagree about supply, lifecycle, evidence, or governance. The second occurs when two implementations encode the same business token using different ledgers, transaction types, or message schemas. A semantic token framework must make $S$ explicit, reusable, and testable without collapsing it into any one $I$.

Table~\ref{tab:reqs} derives design requirements from recurring failure modes.

\begin{table*}[!t]
\caption{Requirements for an Implementation-Neutral Token Semantics Layer}
\label{tab:reqs}
\centering
\footnotesize
\begin{tabular}{@{}p{0.09\textwidth}p{0.25\textwidth}p{0.30\textwidth}p{0.24\textwidth}@{}}
\toprule
\textbf{Req.} & \textbf{Failure mode} & \textbf{Requirement} & \textbf{TTF mechanism} \\
\midrule
R1 & Tokens share an interface but differ in meaning. & Define semantic commitments independently of platform APIs. & Base types, behaviors, property sets, formulas. \\
R2 & Semantically equivalent tokens use different ledgers. & Preserve a platform-neutral definition prior to binding. & Token templates and definitions. \\
R3 & Bridges move balances but lose lifecycle or evidence. & Capture state invariants, lifecycle effects, and required data. & Behavior artifacts, property sets, control messages. \\
R4 & Specifications are prose-only and cannot be validated. & Provide machine-processable metadata and typed composition. & JSON artifacts, formula syntax, object model. \\
R5 & Domain terms fragment across projects. & Reuse governed domain artifacts rather than ad hoc labels. & Versioned artifact libraries. \\
R6 & Implementations satisfy APIs but not token obligations. & Test conformance at message, state, and trace layers. & Platform-neutral control messages. \\
R7 & Artifact evolution breaks deployed meanings. & Pin versions and define compatibility rules. & Artifact version folders and references. \\
R8 & Legal, business, and technical users cannot inspect the same definition. & Support human-readable and machine-readable specification. & Markdown, schemas, formulas, generated documentation. \\
\bottomrule
\end{tabular}
\end{table*}

\section{A Semantic Composition Model for TTF}

\subsection{Artifact Metamodel}

TTF is organized as a repository of artifacts. The taxonomy model describes TTF as a GitHub and file-system based collection of text artifacts, with an object model used by applications \cite{ttf_model}. Artifacts are stored under folders based on type: Base, Behavior, Behavior-Group, Property-Set, and TokenTemplates \cite{ttf_model}. The artifact-format documentation explains that artifact folders may contain JSON definitions, Protocol Buffer files, markdown documentation, sequence diagrams, and custom supporting files \cite{ttf_artifact_format}.

We model a TTF artifact as

\begin{equation}
a = \langle id, ver, kind, sig, req, inv, msg, doc \rangle,
\end{equation}

where $id$ is a stable artifact identifier, $ver$ is a version, $kind \in \Kind$, $sig$ is a typed signature, $req$ is a finite set of requirements over other artifacts or state variables, $inv$ is a finite set of invariants induced by the artifact, $msg$ is a finite set of control-message definitions, and $doc$ is the human-readable documentation. The kind universe is

\begin{equation}
\begin{aligned}
\Kind = \{&\Base, \Beh, \BGroup,\\
&\PSet, \Msg\}.
\end{aligned}
\end{equation}

A behavior group is interpreted as a named expansion into behavior artifacts:

\begin{equation}
expand(g) = \{h_1,\ldots,h_n\}, \quad g \in \BGroup.
\end{equation}

A property set contributes typed state variables and data constraints. A behavior contributes transitions, preconditions, postconditions, and possible control messages. A base contributes representation, unit, value, and identity assumptions. Table~\ref{tab:artifacts2} summarizes the principal artifact classes.

\begin{table}[!t]
\caption{Core TTF Artifact Classes in the Formal Model}
\label{tab:artifacts2}
\centering
\footnotesize
\begin{tabular}{@{}p{0.25\columnwidth}p{0.61\columnwidth}@{}}
\toprule
\textbf{Artifact class} & \textbf{Semantic role} \\
\midrule
Base token type & Introduces identity and representation assumptions, such as fungible, non-fungible, whole, fractional, or singleton forms. \\
Behavior & Introduces capabilities, constraints, transitions, messages, or lifecycle effects, such as transferable, mintable, burnable, delegable, or role-controlled. \\
Behavior group & Expands to a reusable package of related behaviors, such as supply-control capabilities. \\
Property set & Introduces typed class-level or instance-level data, off-chain references, evidence fields, or domain attributes. \\
Token template & Composes base, behavior, behavior group, and property-set artifacts into a reusable design. \\
Token definition & Binds a formula to concrete metadata, artifact versions, domain values, and conformance expectations. \\
Control message & Defines implementation-neutral invocation, response, state, or property messages used for certification and testing. \\
\bottomrule
\end{tabular}
\end{table}

\subsection{Formula Syntax}

The compact representation of a TTF design is the token formula. The visual TTF syntax uses symbols such as $\tau_F$ for fungible and $\tau_N$ for non-fungible, while tooling representations use strings such as \texttt{tF} and \texttt{tN} \cite{ttf_taxonomy}. A simplified grammar is shown in Table~\ref{tab:grammar}. The grammar is intentionally abstract; specific artifact libraries instantiate terminals and symbols.

\begin{table}[!t]
\caption{Abstract Grammar for Token Formulas}
\label{tab:grammar}
\centering
\footnotesize
\begin{tabular}{@{}p{0.25\columnwidth}p{0.61\columnwidth}@{}}
\toprule
\textbf{Construct} & \textbf{Grammar} \\
\midrule
Formula & $F ::= [B\{A\} + P + C]$ \\
Base & $B ::= b$ where $b$ is a base artifact \\
Artifacts & $A ::= \epsilon \mid a \mid a,A$ where $a$ is a behavior or behavior group \\
Property sets & $P ::= \epsilon \mid p \mid p+P$ where $p$ is a property-set artifact \\
Children & $C ::= \epsilon \mid F \mid F+C$ \\
Reference & $r ::= id \mid id@ver$ \\
\bottomrule
\end{tabular}
\end{table}

A normalized formula first expands behavior groups, resolves versions, and orders artifact identifiers canonically:

\begin{equation}
normalize(F) = \langle B, H, P, C, V \rangle,
\end{equation}

where $B$ is exactly one base, $H$ is the expanded set of behavior artifacts, $P$ is the set of property sets, $C$ is the set of child formulas, and $V$ records version bindings.

A full token definition is then

\begin{equation}
D_F = \langle B, U, V_t, R, S, H, P, C, M, G \rangle,
\end{equation}

where $U$ is token unit, $V_t$ is value type, $R$ is representation type, $S$ is supply model, $M$ is the set of control messages, and $G$ is governance metadata such as issuer, artifact authority, version policy, and deprecation status.

\subsection{Semantic Interpretation}

The semantic interpretation of a formula is a tuple

\begin{equation}
\sem{F} = \langle \Sigma_F, \mathcal{M}_F, \mathcal{I}_F, \mathcal{T}_F, \mathcal{E}_F \rangle,
\end{equation}

where $\Sigma_F$ is the admissible state space, $\mathcal{M}_F$ is the set of required messages and observations, $\mathcal{I}_F$ is the set of invariants, $\mathcal{T}_F \subseteq \Sigma_F \times \mathcal{M}_F \times \Sigma_F$ is the transition relation induced by selected behaviors, and $\mathcal{E}_F$ is the set of externally observable events, receipts, or evidence records.

A property set $p$ contributes state variables and constraints:

\begin{equation}
\sem{p} = \langle fields(p), constraints(p) \rangle.
\end{equation}

A behavior $h$ contributes transition structure:

\begin{equation}
\sem{h} = \langle msg(h), pre(h), post(h), err(h), event(h) \rangle.
\end{equation}

The semantics of a formula is composition over its selected artifacts:

\begin{align}
\Sigma_F &= state(B) \times \prod_{p \in P} fields(p), \\
\mathcal{I}_F &= inv(B) \cup \bigcup_{h \in H} inv(h) \cup \bigcup_{p \in P} constraints(p), \\
\mathcal{M}_F &= \bigcup_{h \in H} msg(h) \cup \bigcup_{p \in P} msg(p), \\
\mathcal{T}_F &= compose\big(\{\sem{h} \mid h \in H\}\big).
\end{align}

This formulation makes explicit a point that interface standards frequently hide: token state is not merely balances or ownership. It may also include evidence pointers, issuer attributes, inspection data, role assignments, lifecycle status, methodology references, retirement status, and domain-specific constraints.

\subsection{Well-Formedness}

A formula is well formed with respect to an artifact library $L$ if all of the following hold.

\begin{enumerate}
\item \emph{Reference validity}: every referenced artifact identifier and version exists in $L$.
\item \emph{Base uniqueness}: exactly one base artifact is selected for each token formula.
\item \emph{Kind correctness}: only behavior artifacts or behavior groups appear inside the behavior set, and only property sets appear in the property-set position.
\item \emph{Dependency satisfaction}: if an artifact requires another artifact, state variable, role, or property set, the requirement is satisfied by the formula or its context.
\item \emph{Compatibility}: no pair of selected artifacts introduces inconsistent invariants or contradictory lifecycle effects.
\item \emph{Version determinacy}: every artifact reference is either pinned to a version or resolved by an explicit version policy.
\item \emph{Child consistency}: child formulas do not violate the parent formula's supply, containment, lifecycle, or governance constraints.
\end{enumerate}

For example, a singleton token with a divisibility behavior is not well formed if singleton implies indivisible quantity. A capped-supply token with an unconstrained minting behavior is not well formed unless the mint transition preserves the cap invariant. A document token with a file property set is incomplete if the property set requires a file hash or immutable URI and no corresponding field is bound.

\subsection{Equivalence and Refinement}

Two formulas are syntactically equivalent if their normalized artifact tuples are identical:

\begin{equation}
F_1 \equiv_{syn} F_2 \iff normalize(F_1) = normalize(F_2).
\end{equation}

They are semantically equivalent under abstraction $\pi$ if their projected state spaces, messages, invariants, and traces agree:

\begin{equation}
F_1 \equiv_{\pi} F_2 \iff \pi(\sem{F_1}) = \pi(\sem{F_2}).
\end{equation}

Refinement is the relation needed for domain extension. Formula $F_2$ refines $F_1$ if it preserves all obligations of $F_1$ while adding constraints, fields, or behaviors that do not violate inherited invariants:

\begin{equation}
F_2 \sqsubseteq F_1 \iff \mathcal{I}_{F_1} \subseteq \mathcal{I}_{F_2} \wedge \Trace(F_2) \subseteq \Trace(F_1),
\end{equation}

where traces are compared at the abstraction level of the parent definition. This allows, for example, a generic transferable document token to be refined into a regulated document token with role-gated transfer while still preserving the parent document-token obligations.

\section{Control Messages and Layered Conformance}

\subsection{Control-Message Semantics}

Control messages are the mechanism by which TTF moves from descriptive taxonomy toward testable interaction. The TTF control-message documentation states that the taxonomy uses Protocol Buffers because they are language- and platform-neutral, and that these messages are used to certify implementations of specifications \cite{ttf_control}. The documentation also explains that transports are not defined by the standard, although gRPC and AMQP are recommended options depending on implementation context \cite{ttf_control}.

We model a control message as

\begin{equation}
m = \langle name, in, out, pre, post, err, event \rangle,
\end{equation}

where $in$ and $out$ are typed schemas, $pre$ is the precondition over the current token state and caller context, $post$ is the required relation between pre-state and post-state, $err$ is the set of expected error cases, and $event$ is the observable evidence emitted by a successful or failed invocation.

For example, the transferable behavior can be associated with request/response messages. The TTF documentation shows a \texttt{TransferRequest} containing a message header, token identifier, destination account identifier, and quantity, and a corresponding \texttt{TransferResponse} containing confirmation data \cite{ttf_control}. The message pair is not a Solidity ABI and not a blockchain transaction format. It is an implementation-neutral description of observable token behavior.

\subsection{Layered Conformance}

An implementation may satisfy a platform ABI while failing to preserve a token's semantic commitments. Therefore, conformance must be layered. Table~\ref{tab:conformance} gives the proposed model.

\begin{table*}[!t]
\caption{Layered Conformance Model for TTF Token Definitions}
\label{tab:conformance}
\centering
\footnotesize
\begin{tabular}{p{0.15\textwidth}p{0.34\textwidth}p{0.39\textwidth}}
\toprule
\textbf{Layer} & \textbf{Conformance question} & \textbf{Evidence or check} \\
\midrule
Formula & Does the formula parse and resolve to known artifact versions? & Grammar check, artifact lookup, version-resolution policy. \\
Artifact & Are selected bases, behaviors, groups, and property sets mutually compatible? & Type checking, dependency checks, invariant satisfiability, conflict detection. \\
Message & Does the implementation realize every required control message? & Adapter mapping from platform calls, commands, or transactions to message schemas. \\
State & Do all reachable states satisfy declared token invariants? & Supply, divisibility, ownership, lifecycle, role, evidence, and property constraints. \\
Trace & Do observable executions satisfy permitted lifecycle and authorization traces? & Projected traces accepted by the transition system induced by $\sem{F}$. \\
Governance & Are artifact versions, issuers, authorities, and deprecation policies respected? & Version pinning, artifact signatures, provenance, registry entries, migration rules. \\
\bottomrule
\end{tabular}
\end{table*}

Let $P$ be a platform implementation and $\alpha_P$ be an adapter from platform-native observations to TTF messages and state projections. $P$ conforms to formula $F$ with respect to adapter $\alpha_P$ if:

\begin{enumerate}
\item $F$ is well formed;
\item for every required message $m \in \mathcal{M}_F$, $\alpha_P$ maps at least one platform operation or query to $m$;
\item every projected reachable state $\alpha_P(s)$ satisfies $\mathcal{I}_F$; and
\item every projected execution trace $\alpha_P(\tau)$ is accepted by the transition relation $\mathcal{T}_F$.
\end{enumerate}

This definition does not require every platform to expose the same API. A Corda state transition, a Fabric chaincode transaction, a Solidity function, and a DAML exercise can all implement the same TTF behavior through different native mechanisms. Conformance is checked at the semantic boundary.

\begin{figure}[!t]
\centering
\begin{tikzpicture}[
  layer/.style={draw, rounded corners, align=center, minimum height=0.55cm, text width=6.0cm, font=\footnotesize},
  arr/.style={-{Latex[length=2mm]}, thick}
]
\node[layer] (f) {Formula conformance};
\node[layer, below=0.25cm of f] (a) {Artifact conformance};
\node[layer, below=0.25cm of a] (m) {Message conformance};
\node[layer, below=0.25cm of m] (s) {State conformance};
\node[layer, below=0.25cm of s] (t) {Trace conformance};
\draw[arr] (f) -- (a);
\draw[arr] (a) -- (m);
\draw[arr] (m) -- (s);
\draw[arr] (s) -- (t);
\end{tikzpicture}
\caption{Conformance stack. Lower layers depend on the syntactic and semantic obligations established by higher layers.}
\label{fig:conformance-stack}
\end{figure}
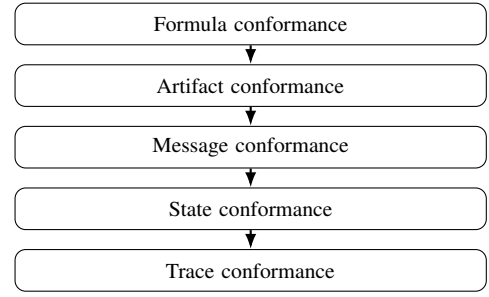

\section{Static Validation and Test Generation}

\subsection{Reference Validation Procedure}

A TTF validator checks token definitions before implementation and before certification. Algorithm~\ref{alg:validate} gives a reference procedure. It is intentionally platform neutral: it does not inspect Solidity, chaincode, or Corda contracts. Instead, it validates the semantic definition that implementations must later realize.

\begin{algorithm}[!t]
\caption{Reference validation for a TTF formula}
\label{alg:validate}
\footnotesize
\begin{algorithmic}[1]
\REQUIRE Formula $F$, artifact library $L$, version policy $\nu$
\ENSURE \textsc{Valid} or set of diagnostic errors $E$
\STATE $E \leftarrow \varnothing$
\STATE Parse $F$ using the formula grammar
\IF{parse fails} \STATE add syntax error to $E$; \RETURN $E$ \ENDIF
\STATE Resolve every artifact reference using $L$ and $\nu$
\FORALL{unresolved or ambiguous references}
  \STATE add reference diagnostic to $E$
\ENDFOR
\STATE Expand behavior groups and normalize artifact order
\IF{formula does not contain exactly one base}
  \STATE add base-uniqueness diagnostic to $E$
\ENDIF
\FORALL{artifacts $a$ in $F$}
  \STATE check kind correctness and dependency satisfaction
  \STATE add missing dependency diagnostics to $E$
\ENDFOR
\FORALL{artifact pairs $(a_i,a_j)$ in $F$}
  \STATE check compatibility of invariants and lifecycle effects
  \STATE add conflict diagnostics to $E$
\ENDFOR
\STATE Construct $\Sigma_F$, $\mathcal{M}_F$, $\mathcal{I}_F$, and $\mathcal{T}_F$
\STATE Check required messages against behavior and property obligations
\STATE Check child formulas against parent supply and lifecycle constraints
\IF{$E=\varnothing$} \RETURN \textsc{Valid} \ELSE \RETURN $E$ \ENDIF
\end{algorithmic}
\end{algorithm}

Validation is not a substitute for security auditing. It does not prove that an implementation is free of reentrancy, arithmetic, consensus, or access-control defects. It prevents a different class of failures: semantic drift between business intent and platform code.

\subsection{Invalid Mutants}

A useful evaluation technique is to construct invalid mutants: formulas that appear plausible but violate semantic constraints. Table~\ref{tab:mutants} gives representative mutants and the diagnostics that a validator should produce.

\begin{table*}[!t]
\caption{Representative Invalid Compositions and Expected Diagnostics}
\label{tab:mutants}
\centering
\footnotesize
\begin{tabular}{p{0.28\textwidth}p{0.34\textwidth}p{0.28\textwidth}}
\toprule
\textbf{Invalid composition} & \textbf{Semantic defect} & \textbf{Expected diagnostic} \\
\midrule
Singleton base with divisibility behavior & Singleton quantity conflicts with subdividable unit semantics. & Unit/behavior incompatibility. \\
Capped supply with unconstrained mint & Mint transition can violate the supply cap. & Missing postcondition preserving cap invariant. \\
Transferable token without ownership or holder state & Transfer cannot be specified without a source and destination ownership relation. & Missing state dependency. \\
Burn behavior with no lifecycle interpretation & Burn is ambiguous among destruction, cancellation, redemption, and retirement. & Missing lifecycle effect or domain refinement. \\
File property set without URI/hash mutability policy & File reference does not define identity, integrity, or update semantics. & Missing required field or property constraint. \\
Carbon claim without methodology and verifier evidence & Environmental claim lacks measurement basis and attestation semantics. & Missing domain property sets. \\
Child token created outside parent supply policy & Containment or issuance violates the parent token definition. & Parent/child consistency error. \\
\bottomrule
\end{tabular}
\end{table*}

These diagnostics are not merely engineering conveniences. They are the mechanism by which TTF raises the quality of token design: mistakes become visible at the specification layer, before the ambiguity is compiled into contracts, schemas, or operational procedures.

\subsection{Generating Conformance Obligations}

Each behavior message can be translated into one or more conformance obligations. A simple obligation has the form

\begin{equation}
\langle setup, request, pre, expected, forbidden \rangle,
\end{equation}

where $setup$ creates a valid initial state, $request$ invokes a control message, $pre$ defines the required precondition, $expected$ defines post-state and event obligations, and $forbidden$ defines states or events that must not occur.

For a transferable token, a generated test may state that if account $x$ owns quantity $q$ and transfer quantity $r \le q$ is requested to account $y$, then the post-state reduces $x$ by $r$, increases $y$ by $r$, preserves total supply, and emits a transfer confirmation. For a burn or retirement behavior, the test must additionally check the correct lifecycle effect and evidence record. For a role-gated behavior, the same request must fail when submitted by a caller that lacks the required role.

\section{Implementation Mapping}

TTF is complementary to implementation standards. It should not be treated as an alternative to ERC-20, ERC-721, ERC-1155, or similar interfaces. Rather, it answers a different question. An implementation standard asks how the token is called on a platform. TTF asks what semantic obligations the token carries across platforms.

Table~\ref{tab:mapping} illustrates mapping from selected TTF concepts to common implementation targets. The purpose is not to claim that every mapping is lossless. On the contrary, the value of the table is that it exposes where a target platform naturally supports a semantic feature, where an extension is required, and where governance or off-chain evidence is needed.

\begin{table*}[!t]
\caption{Illustrative Mapping from TTF Concepts to Implementation Targets}
\label{tab:mapping}
\centering
\footnotesize
\begin{tabular}{p{0.15\textwidth}p{0.16\textwidth}p{0.16\textwidth}p{0.18\textwidth}p{0.18\textwidth}}
\toprule
\textbf{TTF concept} & \textbf{ERC-20} & \textbf{ERC-721} & \textbf{Hyperledger Fabric} & \textbf{Corda/DAML-style shared state} \\
\midrule
Token identity & Contract-level class; balances by address. & Token identifier plus contract address. & World-state key and chaincode namespace. & State reference, contract type, or template instance. \\
Ownership/holding & Balance mapping. & Owner relation for token identifier. & State field or key ownership relation. & Participants, owner field, or parties in state/template. \\
Transfer & \texttt{transfer} and \texttt{transferFrom}. & \texttt{transferFrom} or safe transfer. & Chaincode transaction updating world state. & Command/exercise creating successor state. \\
Supply constraint & Contract variable plus mint/burn logic. & Token creation and burn policy. & Chaincode invariant over state. & Contract invariant over state transitions. \\
Property set & Usually extension or external metadata. & Metadata URI plus custom fields. & JSON/protobuf fields in world state. & State/template fields. \\
Lifecycle status & Custom extension. & Custom extension. & Explicit state field and transaction type. & Contract state and command semantics. \\
Evidence reference & URI/hash extension or event. & Metadata URI/hash extension. & On-chain field plus off-chain evidence store. & State field plus attachment or external reference. \\
Control message & Adapter over function calls. & Adapter over function calls. & Adapter over transaction proposal and response. & Adapter over command/exercise and resulting state. \\
\bottomrule
\end{tabular}
\end{table*}

The mapping problem is therefore not a universal code-generation problem. It is a traceability problem. For each TTF artifact, the implementer should state how the target realizes the artifact, which obligations are native, which require custom code, and which are discharged by external governance or evidence systems.

\section{Case Studies}

\subsection{Document Token}

A document token is often described casually as an NFT, but that label is insufficient. A document token may require singleton quantity, indivisibility, issuer identity, file properties, access constraints, transfer behavior, burn or cancellation behavior, provenance, and audit receipts. The IWA fact card gives the example formula

\begin{equation}
\texttt{[tN\{\textasciitilde d,t,s,e,b\}+phFile]}
\end{equation}

for a document token and explains that TTF combines token bases, behaviors, and property sets into token formulas \cite{ttf_fact_card}.

A stronger TTF definition expands this compact formula into obligations. The base establishes non-fungible identity. The indivisibility behavior prevents partial transfer. Transfer behavior defines ownership change. Burn or cancellation behavior must specify whether the document token is invalidated, archived, retired, or merely removed from circulation. The file property set must define at least a file URI or storage reference, integrity hash, mutability policy, and issuer.

The implementation mapping then becomes auditable. In ERC-721, file properties may be encoded through metadata URI and custom fields; in Fabric, they may be stored as world-state fields; in Corda, they may be part of the state object. The semantic obligation is the same: the implementation must preserve document identity, file-reference integrity, transfer rules, lifecycle status, and evidence receipts.

\subsection{Warehouse Receipt Token}

A warehouse receipt token represents a claim associated with stored goods. Interface labels are especially weak here. A transfer function does not specify quantity, grade, storage location, inspection record, issuer authority, expiration, encumbrance, or redemption. An illustrative formula is

\begin{equation}
\begin{aligned}
&\text{\texttt{[tN\{\textasciitilde d,t,e,b,r,enc\}}}\\
&\text{\texttt{+psReceipt+psInspection]}}.
\end{aligned}
\end{equation}

Here \texttt{psReceipt} contains issuer, goods description, quantity, unit, location, expiration, and holder fields; \texttt{psInspection} contains quality grade, inspection time, inspector identity, and evidence reference; \texttt{r} denotes a redeemable lifecycle behavior; and \texttt{enc} denotes encumbrance or pledge behavior.

The principal invariants are: the receipt has a single current holder; redeemed receipts cannot be transferred; encumbered receipts cannot be transferred except under the encumbrance policy; quantity and unit must match the stored-goods claim; and inspection evidence must remain linked to the receipt. These invariants are not captured by an ERC-721 identifier alone. They are token-level semantics.

\subsection{Carbon and Digital MRV Tokens}

Sustainability use cases show both the power and the boundary of TTF. Carbon emissions, carbon removal, renewable energy, biodiversity, and other ecological instruments are not simple balances. They involve measurement, reporting, verification, quality standards, evidence packages, roles, origination processes, issuance, aggregation, retirement, and claims.

The Carbon Emission Token Protocol uses TTF to define and guide emissions tokenization \cite{gbbc_cet}. The dMRV specification states that TTF focuses on defining individual tokens, while the dMRV artifact complements token definitions by capturing relationships among sets of tokens and broader processes \cite{dmrv_spec}. That distinction is architecturally important. TTF defines token-level semantics: base type, behaviors, data properties, formulas, and control messages. The dMRV process layer defines how multiple tokens, data objects, roles, agreements, and checkpoints interact over time.

An illustrative carbon or dMRV token formula is

\begin{equation}
\begin{aligned}
&\text{\texttt{[tF\{d,t,b,SC,retire\}}}\\
&\text{\texttt{+psMRV+psMethod+psVerify]}}.
\end{aligned}
\end{equation}

The formula alone is not enough; the expanded definition must distinguish issuance from measurement, verification from attestation, transfer from retirement, and retirement from ordinary burn. A retirement operation should not be treated as simple destruction of a balance. It must create durable evidence that a claim has been consumed for a specific purpose and should not be double counted.

\subsection{Cross-Case Summary}

Table~\ref{tab:cases} summarizes the three case studies. The important column is not the formula itself, but the semantic obligations made visible by the formula and artifact expansion.

\begin{table*}[!t]
\caption{Case-Study Analysis of Semantic Obligations}
\label{tab:cases}
\centering
\footnotesize
\begin{tabular}{@{}p{0.14\textwidth}p{0.20\textwidth}p{0.28\textwidth}p{0.24\textwidth}@{}}
\toprule
\textbf{Case} & \textbf{Formula pattern} & \textbf{Required semantic obligations} & \textbf{Interface-only gap} \\
\midrule
Document token & Non-fungible, indivisible, file property set, transfer, burn/cancel. & File identity, hash or URI policy, issuer, lifecycle status, transfer authorization, audit receipt. & NFT interface does not distinguish document identity, mutable file references, or cancellation semantics. \\
Warehouse receipt & Non-fungible receipt with inspection, encumbrance, transfer, redemption. & Quantity/unit, holder, issuer, storage location, inspection evidence, expiration, encumbrance policy, redemption finality. & Transfer API does not encode title, inspection, pledge, or redemption obligations. \\
Carbon/dMRV token & Fungible or unitized claim with MRV, methodology, verification, supply control, retirement. & Measurement boundary, methodology, verifier, evidence package, issuance authority, aggregation, retirement evidence. & Balance and burn semantics do not distinguish retirement, cancellation, and double-count prevention. \\
\bottomrule
\end{tabular}
\end{table*}

\section{Analytical Evaluation}

The evaluation question is not whether TTF replaces smart-contract verification or platform standards. It does not. The question is whether the semantic composition model exposes obligations that are otherwise absent or ambiguous at the interface layer. The answer from the case studies is affirmative across three dimensions.

\subsection{Semantic Coverage}

For each case, the formula expands beyond interface methods into representation, behavior, and data obligations. The document token requires file identity and lifecycle semantics. The warehouse receipt requires goods, location, inspection, encumbrance, and redemption semantics. The carbon/dMRV token requires methodology, verifier, evidence, issuance, and retirement semantics. These obligations are not merely metadata; they affect permitted transitions and conformance tests.

\subsection{Error Detection}

The invalid mutants in Table~\ref{tab:mutants} show errors that an interface checker would typically miss. An ERC-compatible transfer function can exist even when the token definition fails to specify ownership state, transfer restrictions, or lifecycle finality. A burn function can exist even when the domain semantics require retirement evidence rather than balance destruction. A metadata URI can exist even when the file-reference policy does not define mutability or integrity.

\subsection{Mapping Transparency}

Table~\ref{tab:mapping} shows that TTF can distinguish native, extended, and externalized obligations across platforms. Some obligations map naturally to platform constructs, such as balances in ERC-20 or state fields in Fabric. Others require custom extensions, such as lifecycle status in ERC-721 or methodology evidence for carbon claims. Still others require governance or external evidence systems. The value of TTF is that these distinctions are explicit.

\subsection{Evaluation Limits}

This evaluation is analytical rather than a performance benchmark. It does not measure runtime overhead, gas cost, developer productivity, or defect rates in a deployed toolchain. Those are appropriate next-step evaluations once a validator, mapping workbench, and test-generation engine are implemented. The present evaluation establishes the formal object to be measured: token semantics as artifact composition plus conformance obligations.

\section{Governance, Versioning, and Runtime Introspection}

Because TTF treats token semantics as artifacts, governance becomes central. The artifact-format documentation states that artifacts are versioned using version folders and version numbers, with the most recent version stored in a \texttt{latest} folder; references may specify artifact versions or default to \texttt{latest} \cite{ttf_artifact_format}. Versioning is critical because token definitions are not static. Industry use cases evolve, regulatory interpretations change, and new behaviors or property sets emerge.

For production-grade use, a token definition should avoid implicit dependence on \texttt{latest} unless the update policy is explicit. A formula intended for certification should pin artifact versions or bind them through a named compatibility policy. Artifact libraries should include provenance, deprecation notices, migration rules, and compatibility metadata. Domain libraries should be curated so that specialized artifacts for finance, supply chain, identity, sustainability, and intellectual property extend the common semantic base without fragmenting it.

Runtime introspection is the natural operational extension. Deployed implementations should be able to expose a TTF formula identifier, artifact-version bindings, a token-definition hash, and a registry URI. Wallets, bridges, auditors, and marketplaces could then ask not only ``what interface does this token expose?'' but also ``what semantic obligations does this token claim to satisfy?''

\section{Limitations and Threats to Validity}

Several limitations are important.

First, token semantics do not exhaust legal meaning. A TTF definition can encode lifecycle, data, and evidence obligations, but it cannot by itself determine whether an off-chain legal right is valid, enforceable, or recognized in a jurisdiction.

Second, off-chain evidence remains a socio-technical assumption. File hashes, methodology references, inspection reports, and verifier attestations depend on storage, oracle, registry, and governance systems outside the token implementation.

Third, platform mappings can be lossy. A target interface may lack native concepts for lifecycle status, encumbrance, retirement evidence, or role-specific issuance. TTF can expose the loss; it cannot automatically eliminate it.

Fourth, semantic conformance does not replace security analysis. A contract can satisfy a TTF formula while still containing implementation vulnerabilities. Conversely, a formally verified contract can still implement the wrong token semantics if the specification is incomplete.

Fifth, artifact governance is itself a standards problem. Without trusted artifact authorities, version policies, and domain-review processes, semantic libraries can fragment into incompatible dialects.

\section{Conclusion}

Interface standards made token ecosystems usable by standardizing how tokens are called. The next interoperability problem is semantic: how to specify what tokens mean across platforms, domains, and implementations. This paper has presented the Token Taxonomy Framework as a typed semantic composition model for implementation-neutral token specifications. The formal model treats TTF artifacts, formulas, behavior composition, property sets, control messages, invariants, and traces as first-class specification objects.

The central claim is that token meaning should be separated from platform binding without being reduced to prose. By composing versioned artifacts into formulas and definitions, TTF makes token semantics inspectable, reusable, comparable, validatable, and governable. By attaching control messages and conformance obligations, TTF creates a path from business intent to implementation testing without requiring every platform to expose the same interface.

The case studies show why this matters. A document token is not merely an NFT; a warehouse receipt is not merely a transferable identifier; a carbon or dMRV token is not merely a balance that can be burned. Each carries semantic obligations that must be preserved if tokenization is to support enterprise, regulatory, and cross-platform interoperability. TTF provides the vocabulary and artifact architecture for preserving those obligations. Future work should implement industrial-strength validators, generate conformance suites from token definitions, define formal mapping adapters for major platforms, and curate governed domain libraries. The underlying thesis, however, is immediate: token standardization must treat semantics, not only code interfaces, as the object of standardization.

\enlargethispage{3\baselineskip}

\end{document}